\documentclass[utf8]{frontiersFPHY} 

\setcitestyle{square} 
\usepackage{url,lineno,microtype,subcaption}
\usepackage[colorlinks,citecolor=blue,linkcolor=blue,anchorcolor=blue,filecolor=blue,urlcolor=blue]{hyperref}
\usepackage[onehalfspacing]{setspace}

\def\keyFont{\fontsize{8}{11}\helveticabold}
\def\firstAuthorLast{K. Godbey and A.S. Umar} 
\def\Authors{K. Godbey\,$^{1}$ and A.S. Umar\,$^{1,*}$}
\def\Address{$^{1}$Department of Physics and Astronomy, Vanderbilt University, Nashville, TN 37235, USA}
\def\corrAuthor{A.S. Umar}
\def\corrEmail{umar@compsci.cas.vanderbilt.edu}

\begin{document}
\onecolumn
\firstpage{1}

\title[Quasifission dynamics in microscopic theories]
{Quasifission dynamics in microscopic theories} 

\author[\firstAuthorLast ]{\Authors} 
\address{} 
\correspondance{} 

\extraAuth{}

\maketitle

\begin{abstract}
In the search for superheavy elements quasifission reactions represent one of the reaction pathways that curtail the formation of an evaporation residue.
In addition to its importance in these searches quasifission is also an interesting dynamic process that could assist our understanding of many-body dynamical shell effects and energy dissipation thus forming a gateway between deep-inelastic reactions and fission.
This manuscript gives a summary of recent progress in microscopic calculations of quasifission employing time-dependent Hartree-Fock (TDHF) theory and its extensions.
\tiny
 \keyFont{ \section{Keywords:} 
Time-dependent Hartree-Fock, Quasifission, Superheavy elements}
\end{abstract}

\section{Introduction}

The ongoing search for discovering new elements in the superheavy regime is perhaps
the most exciting but at the same time challenging tasks in low-energy nuclear 
physics~\cite{dullmann2015}. These searches were historically motivated by theoretical predictions of an
\textit{island of stability}, somewhat detached from the far end of the chart-of-nuclides~\cite{bender1999,nazarewicz2002, cwiok2005,pei2009a},
due to quantum mechanical shell closures.
The experimental search for the so called superheavy elements (SHE) was initially done
by using target projectile combinations that minimized the excitation energy of compound
nuclei that was formed in reactions studied in the vicinity of the Coulomb barrier. For this
reason these reactions are commonly referred to as \textit{cold fusion} reactions and primarily
involved closed shell nuclei such as $^{208}$Pb target and projectiles in the chromium
to zinc region. The cold fusion experiments were able to produce elements $Z$=107-113~\cite{hofmann2002,munzenberg2015,morita2015},
but showed no indication that extending them to heavier elements were feasible. The identification of a
SHE is done through the decay properties of a formed evaporation residue. In such
reactions involving heavy elements the dominant reaction processes are quasifission (QF)
and fusion-fission (FF), which are expected to strongly suppress the formation of an
evaporation residue at higher excitation energies. For this reason it was a major
surprise to observe that the so called \textit{hot fusion} reactions, despite of their
higher excitation energy, were able to synthesize elements Z=113-118~\cite{oganessian2015,roberto2015}.
The hot fusion reactions utilized actinide targets with $^{48}$Ca projectiles.
To further pursue the hot fusion reactions with heavier projectiles to reach
elements $Z>120$ requires a deeper understanding of the reaction pathways leading
to an evaporation residue, particularly QF and FF components.
In all of these reactions the evaporation residue cross-section
is dramatically reduced due to the quasifission (QF) and fusion-fission (FF) processes.
These processes occur during the reactions of heavy systems and correspond to
excited fission channels in the classically allowed
regime above the barrier and require a combination of statistical 
and truly dynamical approaches which are not necessarily confined to a
collective subspace.
Fusion-fission occurs after the formation of a composite system which then
fissions due to its excitation, ultimately resulting in a fragment distribution
that is peaked at equal mass breakup of the composite system.
Quasifission occurs at a considerably shorter time-scale than fusion-fission~\cite{durietz2011,toke1985,shen1987}
and is characterized by reaction fragments
that differ significantly in mass from the original target/projectile nuclei.
Quasifission for being one of the primary reaction
mechanism that limits the formation of superheavy nuclei~\cite{sahm1984,gaggeler1984,schmidt1991} has been
the subject of intense experimental studies of~\cite{toke1985,shen1987,hinde1992,hinde1995,hinde1996,itkis2004,knyazheva2007,hinde2008,nishio2008,kozulin2014,durietz2011,itkis2011,lin2012,nishio2012,simenel2012b,durietz2013,williams2013,kozulin2014,wakhle2014,hammerton2015,prasad2015,prasad2016}.
Studies have also shown a strong impact of the entrance channel characteristics, including deformation~\cite{hinde1995,back1996,hinde1996,umar2006a,hinde2008,nishio2008,oberacker2014} and shell structure~\cite{simenel2012b} of the reactants.
The final phase of the dynamics is also impacted by the fissility of the composite system~\cite{lin2012,durietz2013}, its neutron richness~\cite{hammerton2015}, and by shell effects in the exit channel~\cite{toke1985,shen1987,itkis2004,nishio2008,morjean2008,fregeau2012,kozulin2010,kozulin2014,wakhle2014}.
A number of theoretical approaches have been developed that describe the quasifission in terms of multi-nucleon transfer (MNT)
processes~\cite{adamian2003,zagrebaev2007,aritomo2009,zhao2016,sekizawa2016,sekizawa2017a,sekizawa2019}.
Recently, time-dependent Hartree-Fock (TDHF) theory have proven to be an excellent tool for studying QF dynamics, and
in particular mass-angle distributions and final fragment total kinetic energies (TKE)~\cite{kedziora2010,wakhle2014,oberacker2014,goddard2015,hammerton2015,umar2015a,umar2015c,umar2016,prasad2016,wang2016,sekizawa2016,godbey2019,sekizawa2017,sekizawa2019b}.
While the fragments produced in TDHF studies are the excited primary fragments~\cite{umar2009a} a number of extensions based on the use of
Langevin dynamics have been successfully applied to de-excite these fragments~\cite{sekizawa2017,umar2017,guo2018d,sekizawa2019b,jiang2020}
Theoretical studies of quasifission dynamics have taught us that dynamics themselves may be dominated by shell effects~\cite{simenel2018,sekizawa2019}.
Despite the apparent strong differences between fission and quasifission, it is interesting to note that
similar shell effects are found in both mechanisms~\cite{godbey2019}. Quasifission
can then potentially be used as an alternative mechanisms to probe fission mode properties. For instance,
this could provide a much cheaper way than fusion-fission to test the influence of $^{208}$Pb shell effects in
super-asymmetric SHE fission.

\section{Microscopic approaches}
The underlying approach to study quasifission on a microscopic basis is the time-dependent Hartree-Fock (TDHF) theory~\cite{negele1982,simenel2012,simenel2018,stevenson2019}.
Alternative approaches employ Langevin dynamics~\cite{zagrebaev2012,karpov2017,saiko2019}.
Indeed, the TDHF calculations of the quasifission process have yielded results that not only
agree with the broad features of the experimental measurements but also shed insight into
the relationship of the data to the properties of the participating nuclei.
Such features include static deformation that induces dependence on the orientation of the nuclei with respect to the 
beam axis, shell effects that can predict the primary fragment charges, as well as the
dependence of quasifission on neutron-rich nuclei.
TDHF calculations give us the most probable reaction outcome for a given set of initial
conditions (e.g. energy, impact parameter, orientation).
However, quantum mechanically a collection of outcomes are possible for each of these initial
conditions. In order to compute such distributions, one must go beyond TDHF and introduce
methods to calculate distribution widths or fluctuations for these reactions.
Much effort has been done to improve the standard mean-field approximation by incorporating the fluctuation mechanism
into the description. At low energies, the mean-field fluctuations make the dominant contribution to the fluctuation
mechanism of the collective motion.
Various extensions have been developed to study the fluctuations of one-body observables.
These include the time-dependent random phase approximation (TDRPA) approach of Balian and V\'en\'eroni~\cite{balian1984,balian1992,broomfield2009,simenel2011,williams2018}, the time-dependent generator coordinate method~\cite{goutte2005}, or the stochastic mean-field (SMF) method~\cite{ayik2008,lacroix2014}.
The effects of two-body dissipation on reactions of heavy systems using the time-dependent density matrix (TDDM)~\cite{tohyama1985,tohyama2002a}
approach have also been recently reported~\cite{assie2009,tohyama2016}.
It is also possible to compute the probability to form a fragment with a given number of nucleons~\cite{koonin1977,simenel2010,sekizawa2013,scamps2013a}, but the resulting fragment mass and charge distributions are often underestimated in dissipative collisions~\cite{dasso1979,simenel2011}.
Recent reviews~\cite{simenel2018,sekizawa2019} succinctly summarize the current state of TDHF (and its extensions) as it has been applied to various MNT reactions.

\section{Insights from TDHF and beyond} 
Experiments to discover new elements are notoriously difficult, with
fusion evaporation residue (ER) cross-sections in pico-barns~(for a recent experimental review see~\cite{vardaci2019}). 
This cross-section is commonly expressed in the product form~\cite{loveland2007}
\begin{equation}
\sigma _\mathrm{ER}=\sum_{L=0}^{J_\mathrm{\max }}\sigma
_\mathrm{cap}(E_\mathrm{c.m.},L)P_\mathrm{CN}(E^*,L) W_\mathrm{sur}(E^*,L)\;,
\label{eq:er}
\end{equation}
where $\sigma _\mathrm{cap}(E_\mathrm{c.m.},L)$ is the capture cross-section at center of mass energy
$E_\mathrm{c.m.}$ and orbital angular momentum $L$. $P_\mathrm{CN}$ is the probability that the composite
system fuses into a compound nucleus (CN) rather than breaking up via quasifission, and
$W_\mathrm{sur}$ is
the survival probability of the fused system against fission.
It is thus clear that to have a good handle on the evaporation residue cross-section estimates
it is important to understand each of these terms as well as possible. In this endeavor both
theory and experiment can have a complementary role. Among these reaction mechanisms quasifission
and fusion-fission can be on the order of millibarns, making it easier to study experimentally. However, the
extraction of the $P_\mathrm{CN}$ requires the proper disentangling of quasifission from 
fusion-fission~\cite{yanez2013,zhu2013,schmitt2019} as it may be given by
\begin{equation}
 P_{CN}=\frac{\sigma_{fusion}}{\sigma_{capture}}=\frac{\sigma_{capture}-\sigma_{quasifission}}{\sigma_{capture}}\;.
 \label{PCN}
\end{equation}
Of these cross-sections fusion-fission arises from an excited and equilibrated composite system and therefore
peaked around equal mass breakup as calculated in a statistical approach~\cite{sahm1984,schmidt1991,back1985,back2014,schmidt2018}.
On the other hand, quasifission, which
is a faster process and thus not fully equilibrated, could also contribute to the equal breakup regime.
Consequently, experimental analysis could use assistance from theory to discern between the two processes.
The capture cross-section, being the sum of quasifission, fusion-fission, and evaporation residue is relatively
easy to measure or calculate and TDHF predictions using the density-constrained TDHF (DC-TDHF) approach have shown to give a relatively
good results~\cite{umar2010a,umar2016}. Below, we discuss various aspects of the progress done in studying
quasifission using TDHF and its extensions.
\begin{figure}[!htb]
        \centering
        \includegraphics*[width=12cm]{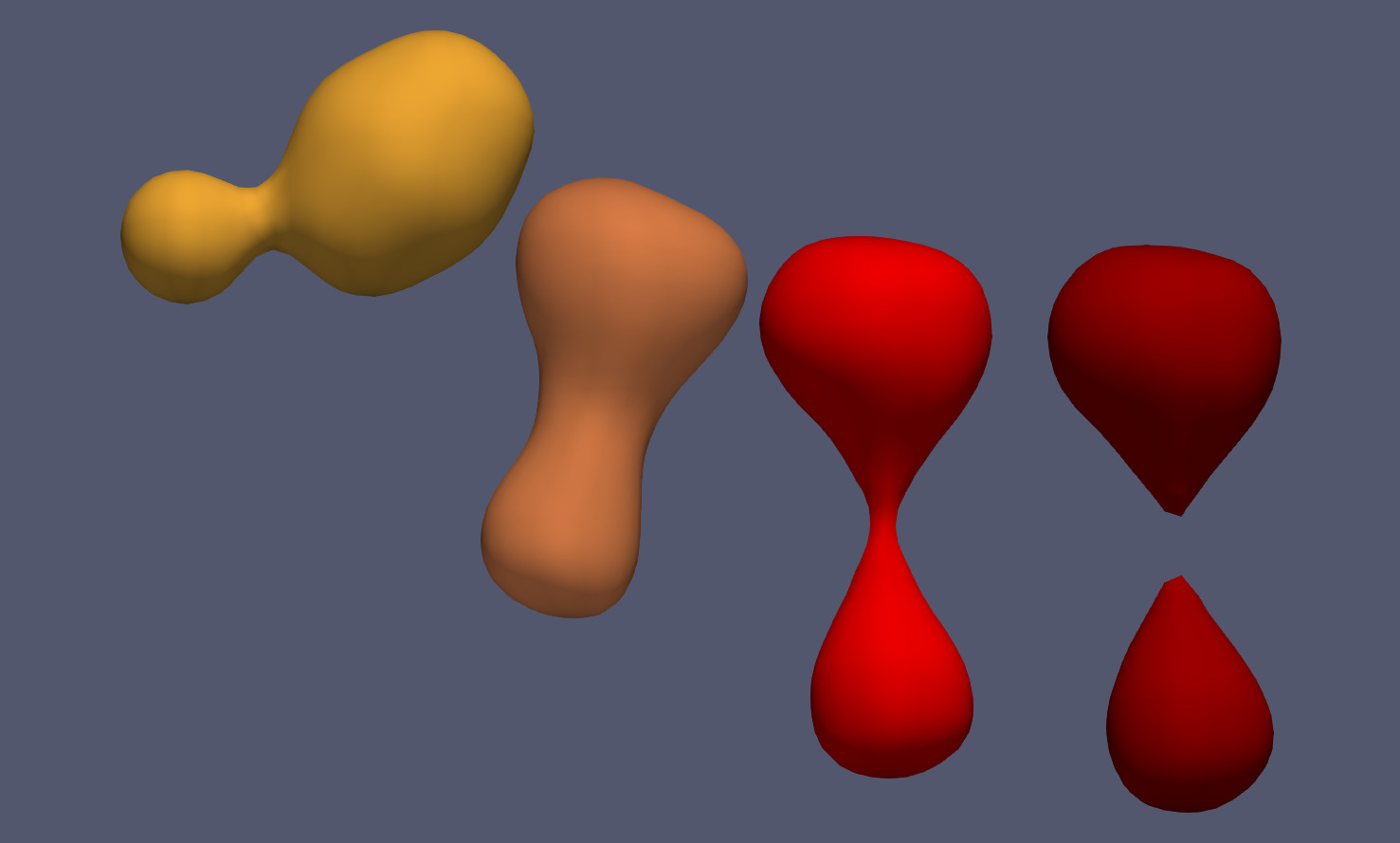}
        \caption{\protect Quasifission in the reaction $^{48}\mathrm{Ca}+{}^{249}\mathrm{Bk}$
at $E_{\mathrm{c.m.}}=234$~MeV and orbital angular momentum $L/\hbar=60$ and the orientation of the
$^{249}\mathrm{Bk}$ with respect to the collision axis $\beta=135^{\circ}$. The darkening of tones depict increasing excitation.}
        \label{fig:evolution}
\end{figure}

\vspace{0.2in}
\subsection{Mass angle distributions}
Study of quasifission together with capture is intimately related to understanding the
process for forming a compound nucleus, the quantity named $P_\mathrm{CN}$ in Eq.~(\ref{eq:er})~\cite{yanez2013}.
Figure~\ref{fig:evolution} shows the time-evolution of the $^{48}\mathrm{Ca}+{}^{249}\mathrm{Bk}$ reaction
at $E_{\mathrm{c.m.}}=234$~MeV and orbital angular momentum $L/\hbar=60$~\cite{godbey2019} and the initial orientation of the
$^{249}\mathrm{Bk}$ with respect to the collision axis $\beta=135^{\circ}$.
For this orbital angular momentum and energy TDHF theory predicts
quasifission.
As the nuclei approach each other, a neck forms between the
two fragments which grows in size as the system begins to rotate.
Due to the Coulomb repulsion and centrifugal forces,
the dinuclear system elongates and forms a very long neck which eventually
ruptures leading to two separated fragments.
In this case the final fragments are $^{203}$Au and $^{94}$Sr.
While the outcome of such reactions in a single TDHF evolution vary greatly depending on the initial conditions, analysis of the fragments' properties can begin to suggest general behavior for systems undergoing quasifission.
For example, the composition of the reaction products can be influenced by shell effects in the outgoing fragments~\cite{godbey2019} which can be inferred by the slight pear shape of the light outgoing fragment at the point of scission in Fig.~\ref{fig:evolution}.

However, the result from a single TDHF trajectory is difficult to extrapolate to the system as a whole so systematic investigations are often performed.
As the reaction products predicted by TDHF give only the most probable outcome for any given collision geometry and energy, quantities like mass angle distributions produced by direct TDHF calculations result in collections of discrete points.
By collecting data from large numbers of TDHF evolutions one can reveal deeper insights into the quasifission process.
Recent studies of the $^{48}$Ca+$^{249}$Bk reaction at $E_\mathrm{c.m.}=234$~MeV with the TDHF approach went beyond solely considering the extreme orientations of the deformed $^{249}$Bk nucleus by undertaking calculations spanning both a range of orientations and a range of angular momenta. The orientation of the deformed
$^{249}$Bk was changed by $15^{\circ}$ steps to cover the full range (0,$\pi$) with orbital angular momentum $L$ changing in units of
$10\hbar$ from $0$ to quasielastic collisions.
A total of 150 TDHF collisions were cataloged and analyzed.
This allows for the study of correlations between, e.g., mass, angle, kinetic energy, as well as to predict distributions of neutron and proton numbers at the mean-field level.
\begin{figure}[!htb]
	\centering
	\includegraphics*[width=8.2cm]{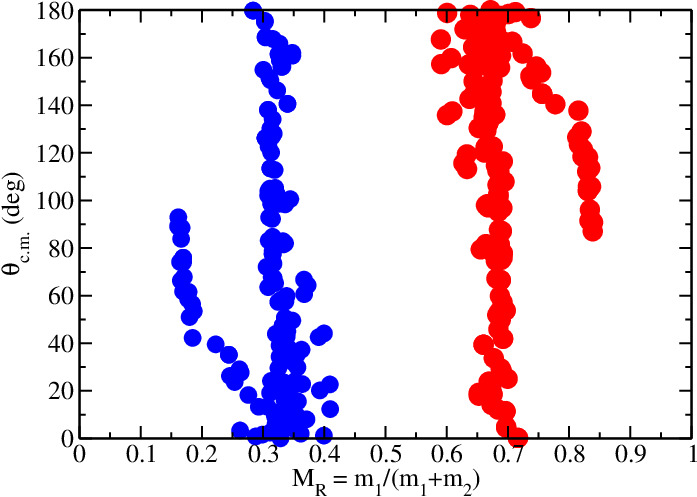}\hspace{0.4in} 
	\includegraphics*[width=8cm]{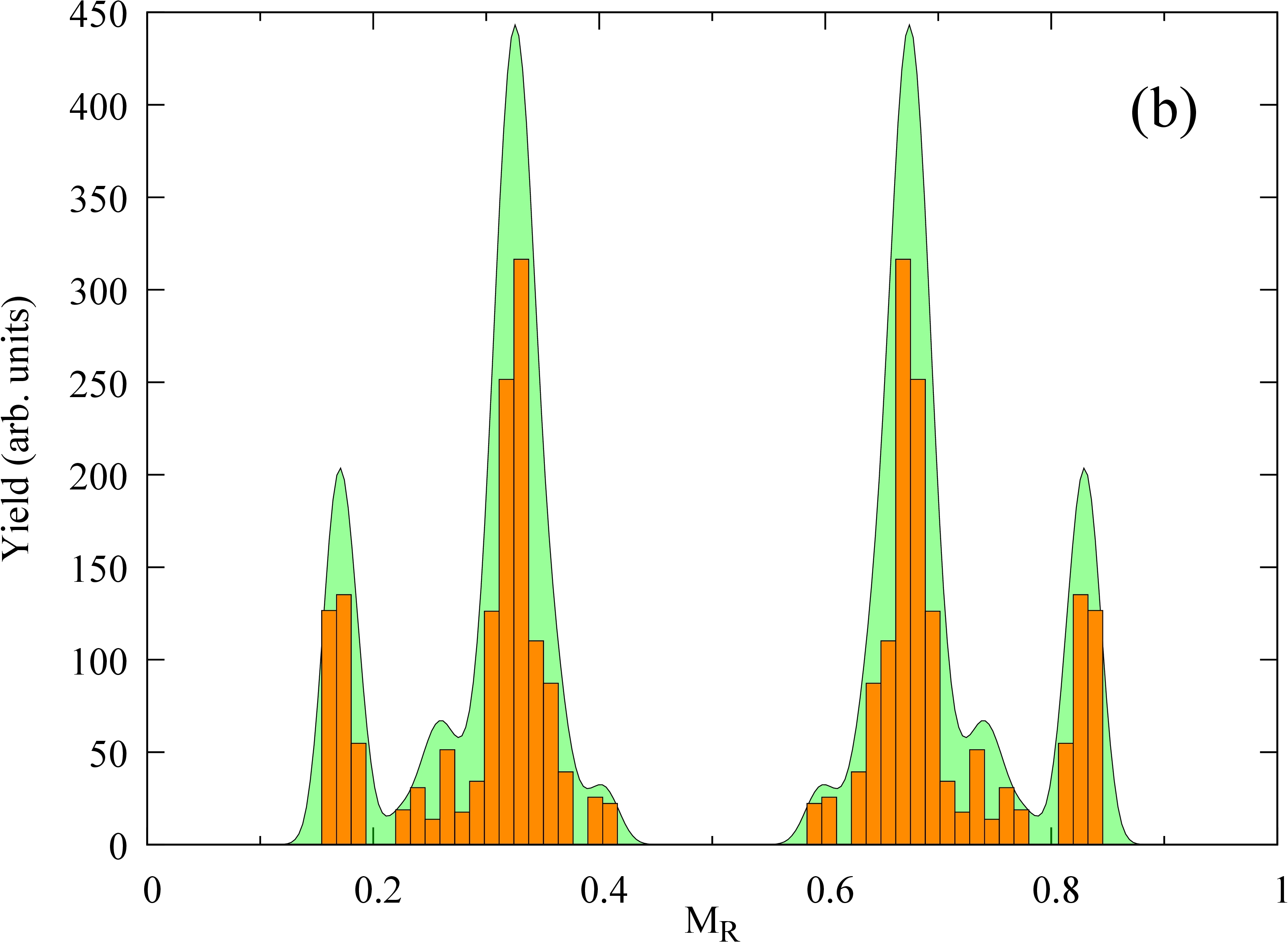} 
	\caption{\protect (a) TDHF MADs for quasifission in the reaction $^{48}\mathrm{Ca}+{}^{249}\mathrm{Bk}$
at $E_{\mathrm{c.m.}}=234$~MeV. (b) The yield (arb. units) as a function of mass ratio $M_\mathrm{R}$ for the same reaction.}
	\label{fig:mad_tdhf}
\end{figure}
In Fig.~\ref{fig:mad_tdhf}(a) we plot the mass angle distribution (MAD) for this reaction. Figure ~\ref{fig:mad_tdhf}(b) shows the corresponding yield in arbitrary units as
a function of the mass ratio $M_\mathrm{R}=M_1/(M_1+M_2)$,
where $M_1$ and $M_2$ are the masses of the final fragments. We note that the yields are strongly peaked at $M_R\sim0.33$ and $0.67$,
with a full width at half maximum FWHM~$\simeq0.1$ corresponding to a standard deviation $\sigma_{M_R}\simeq0.042$.
The purpose of this figure is to compare quantitatively the relative contributions to the yields when going from central to peripheral collisions.
For instance, we see that, because of the $2L+1$ weighting factor, the most central collisions with $L\le20\hbar$, which are found at backward angles, have the smallest contribution to the total yield.
Despite the discrete nature of the data, the tight grouping of points indicates a peak in production probability in certain mass regions which will be discussed further in the next section.

\begin{figure}[!htb]
	\centering
	\includegraphics*[width=16cm]{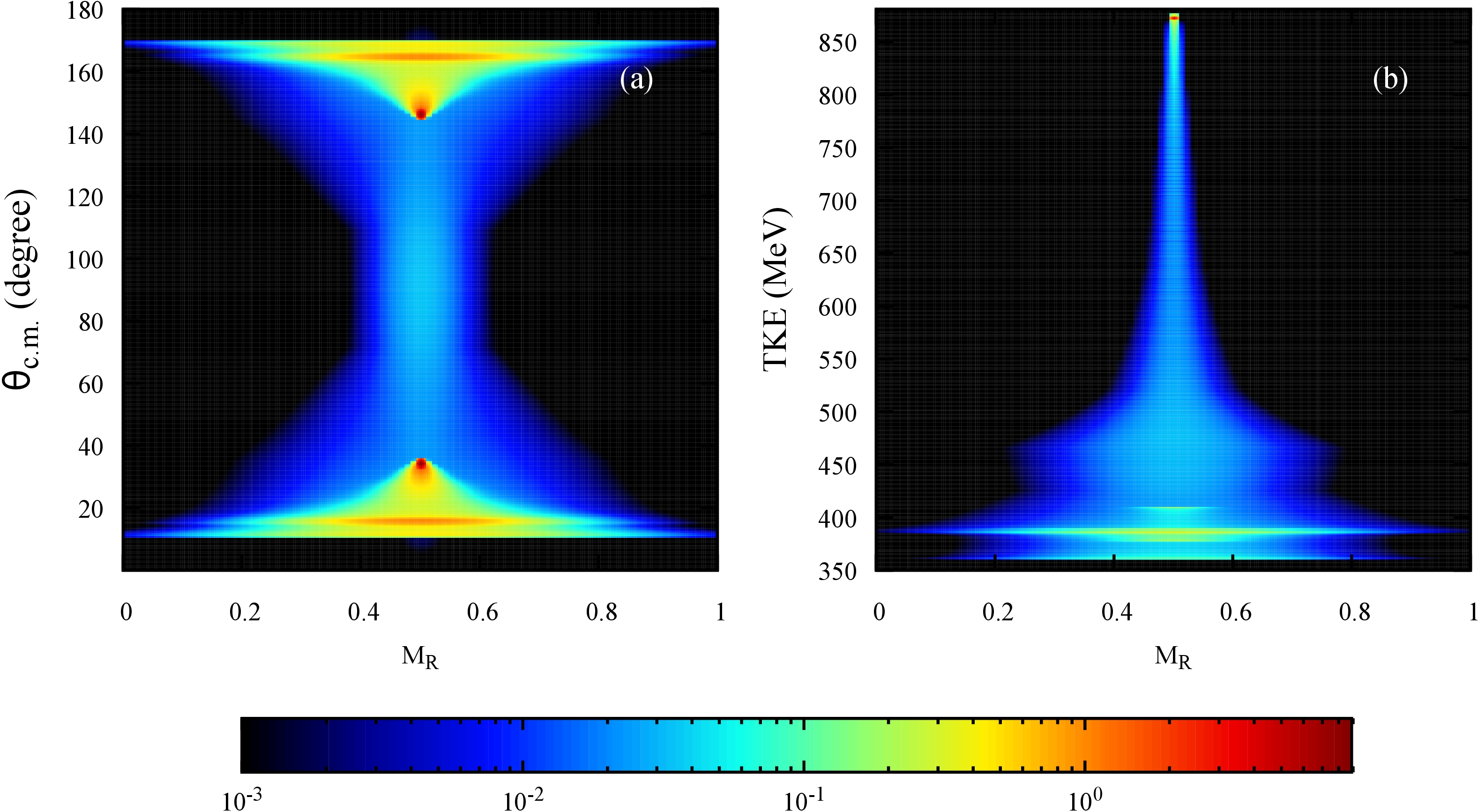}
	\caption{\protect Mass angle (a) and mass energy (b) distributions predicted by TDRPA for the $^{176}\mathrm{Yb}+^{176}\mathrm{Yb}$ collision in the side-side orientation at $E_\mathrm{c.m.}=880$~MeV. Units are in millibarns per degree (a) and millibarns per MeV (b).}
	\label{fig:mad_tdrpa}
\end{figure}

While nucleon transfer fluctuations can be calculated in TDHF, the ability to compare with experiment is still limited by the fact that TDHF vastly under predicts the widths of these distributions.
Ideally, calculations would account for fluctuations in quantities such as particle transfer, scattering angles, and total kinetic energies in the exit channel to more closely obtain what is observed experimentally.
The simplest method for calculating these widths is the particle-number projection for the final fragments~\cite{simenel2010,sekizawa2013,scamps2013a,scamps2017b}.
However, these widths are still seriously underestimated.
This is where extensions such as TDRPA~\cite{balian1984,broomfield2009,simenel2011,williams2018} and SMF~\cite{ayik2008,lacroix2014} have proved to be vital theoretical tools for studying deep inelastic and quasifission reactions as both techniques provide methods to calculate both fluctuations and correlations of neutron and proton transfer based on a TDHF trajectory.
Figure~\ref{fig:mad_tdrpa} shows predicted mass angle and mass energy distributions for the $^{176}\mathrm{Yb}+^{176}\mathrm{Yb}$ system from TDRPA.
Production cross-sections are obtained by integrating the probabilities calculated from the predicted fluctuations over a range of impact parameters.
Such calculations further extend the insight offered by the base TDHF theory and promise to be of great use for designing future MNT experiments.

An alternate approach to TDRPA calculations for beyond the mean-field approximation can be formulated by incorporating the fluctuations in a manner that is consistent with the quantal fluctuation-dissipation relation, namely the SMF method~\cite{lacroix2014}.
In a number of studies it has been demonstrated that the SMF approach improves the description of nuclear collision dynamics by including fluctuation mechanisms of the collective motion.
Most applications have been carried out in collisions where a di-nuclear structure is maintained.
In this case it is possible to define macroscopic variables by a geometric projection procedure with the help of the window dynamics.
The SMF approach gives rise to a Langevin description for the evolution of macroscopic variables.
A limited study for central collisions was published in~\cite{ayik2016}.
A general approach for non-central collisions has been developed~\cite{ayik2018} and used to calculate multi-nucleon transfer and heavy-isotope production in $^{136}\mathrm{Xe}+{}^{208}\mathrm{Pb}$ collisions~\cite{ayik2019,ayik2019b}.

\vspace{0.2in}
\subsection{Deformed shell effects in quasifission}
Returning to the inference of shell effects influencing fragment production, this phenomenon can also be seen through thorough TDHF studies of a particular system and systematically analyzing the fragments produced for different impact parameters and deformation orientations.
TDHF studies of quasifission dynamics have taught us that the dynamics of a system may be dominated
by shell effects~\cite{simenel2018,sekizawa2019}.
An interesting finding of these TDHF studies is the prediction of the role of shell effects
which favor the formation of magic fragments, in particular in the $Z=82$ region in reactions involving an actinide collision partner~\cite{wakhle2014}.
This prediction has been later confirmed experimentally by Morjean and collaborators~\cite{morjean2017}.
In addition, the calculations show that these shell effects strongly depend on the orientation of deformed actinide.
Deformed shell effects in the region of $^{100}$Zr have also been invoked to interpret the outcome of TDHF simulations of $^{40,48}$Ca+$^{238}$U, $^{249}$Bk collisions~\cite{oberacker2014,umar2016}.

\begin{figure}[!htb]
    \begin{center}
            \includegraphics*[width=8cm]{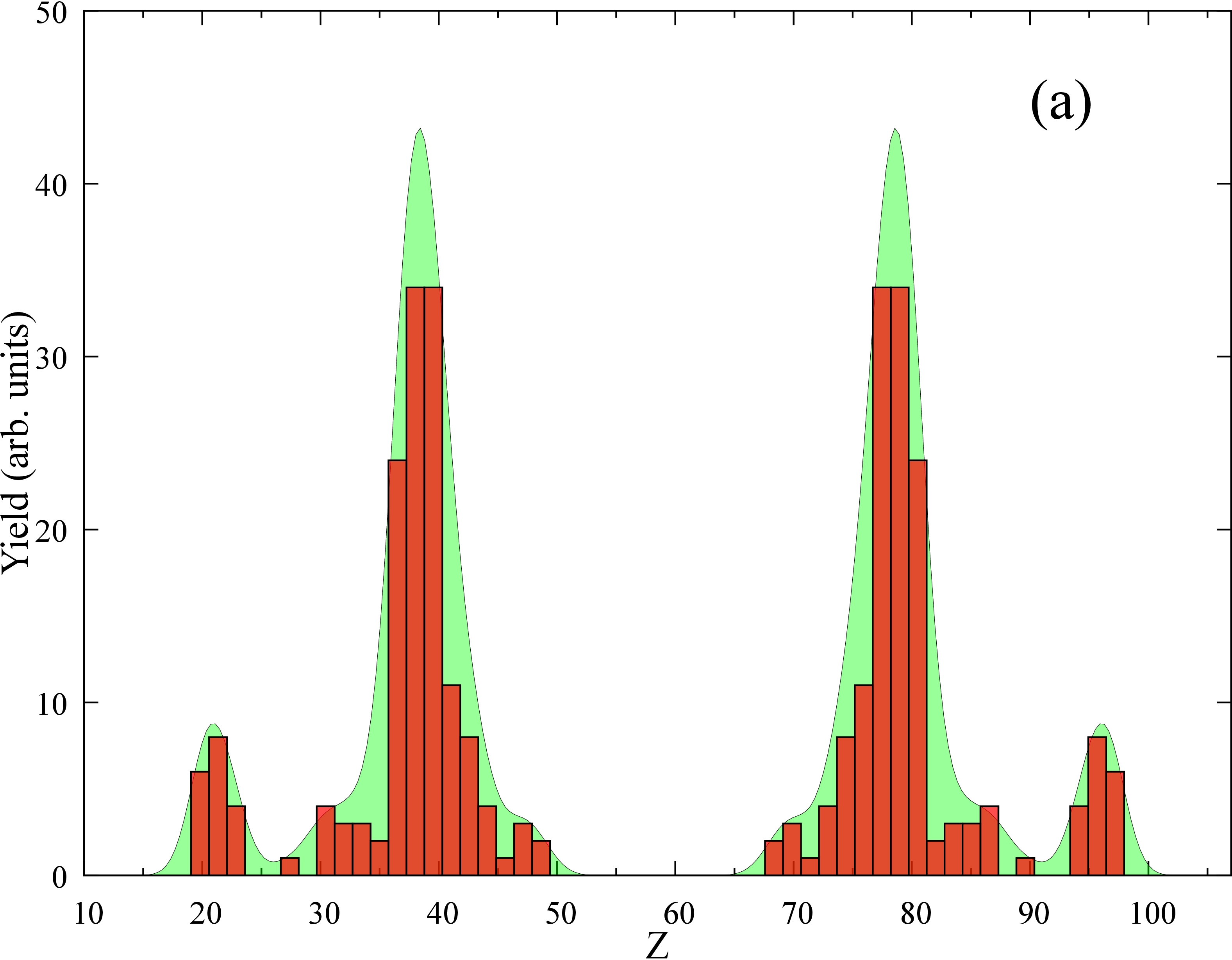}\hspace{0.4in}
            \includegraphics*[width=8.3cm]{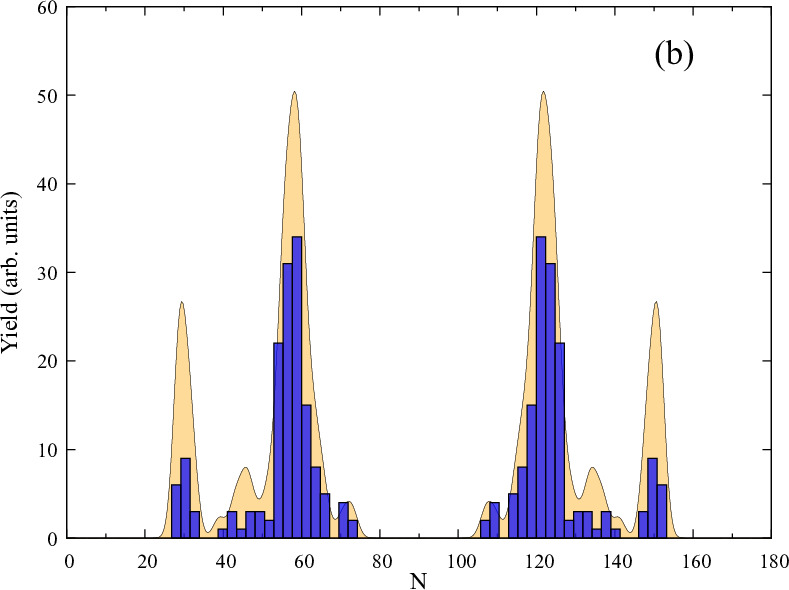}
            \caption{\protect(a) Fragment charge yield (histogram) and (b) Neutron yields for the 
            reaction $^{48}\mathrm{Ca}+{}^{249}\mathrm{Bk}$ at $E_\mathrm{c.m.}=234$~MeV.
            The smooth representations of the histograms are obtained by using a kernel density estimation with bandwidth 0.012.\label{fig:CaBk}}
    \end{center}
\end{figure}

Such results are shown in Fig.~\ref{fig:CaBk} for the reaction $^{48}\mathrm{Ca}+{}^{249}\mathrm{Bk}$ at $E_\mathrm{c.m.}=234$~MeV.
Previous studies of the quasifission dynamics have taught us that dynamics may be dominated
by shell effects~\cite{simenel2018,sekizawa2019,godbey2019}.
These distributions are used to identify potential shell gaps driving quasifission.
In Fig.~\ref{fig:CaBk}(a) we plot the charge yield obtained for this reaction.
The right frame in Fig.~\ref{fig:CaBk}(b) shows the expected neutron yield distributions.
One of the main driving features of this work was to show that
shell effects similar to those observed in fission affect the formation of quasifission fragments. 
For this system the $Z=82$ shell effect does not seem to play a major role contrary to previous TDHF
observations for the Ca+U target projectile combinations.
We also point out
that mass-angle correlations could be used to experimentally isolate the fragments influenced by $N=56$ octupole shell gaps~\cite{scamps2018,scamps2019,godbey2019}.
We also find that more peripheral collisions are centered about the proton number $Z=40$ confirming similar observations
from past calculations~\cite{oberacker2014} that the $^{100}$Zr region plays an important role in determining the lighter fragments due to
the existence of strongly bound highly deformed Zr isotopes in this region~\cite{blazkiewicz2005}.

\vspace{0.2in}
\subsection{Mass equilibration}
\begin{figure}[!htb]
\centering
\includegraphics*[width=8cm]{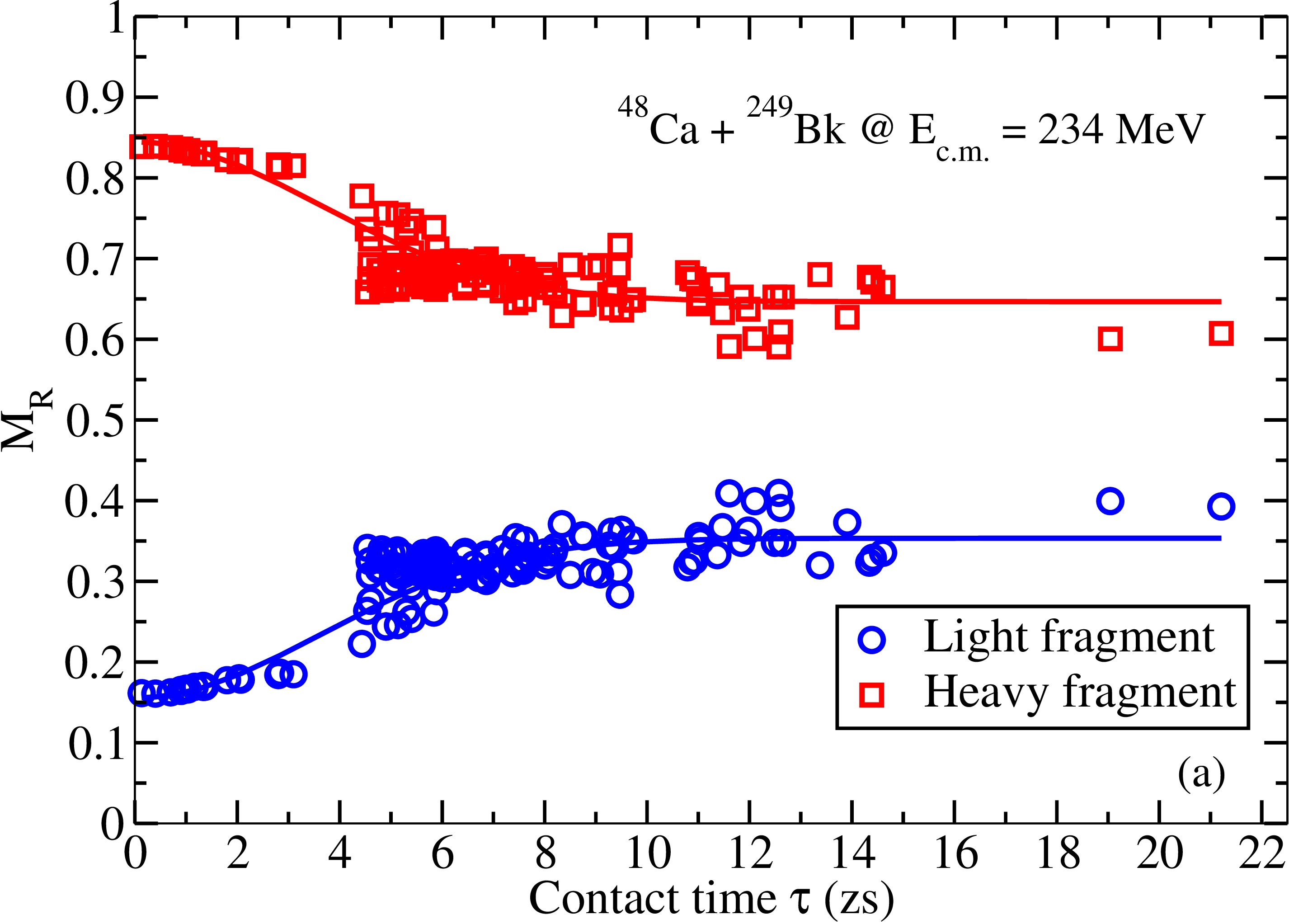}\hspace{0.4in} 
\includegraphics*[width=8.1cm]{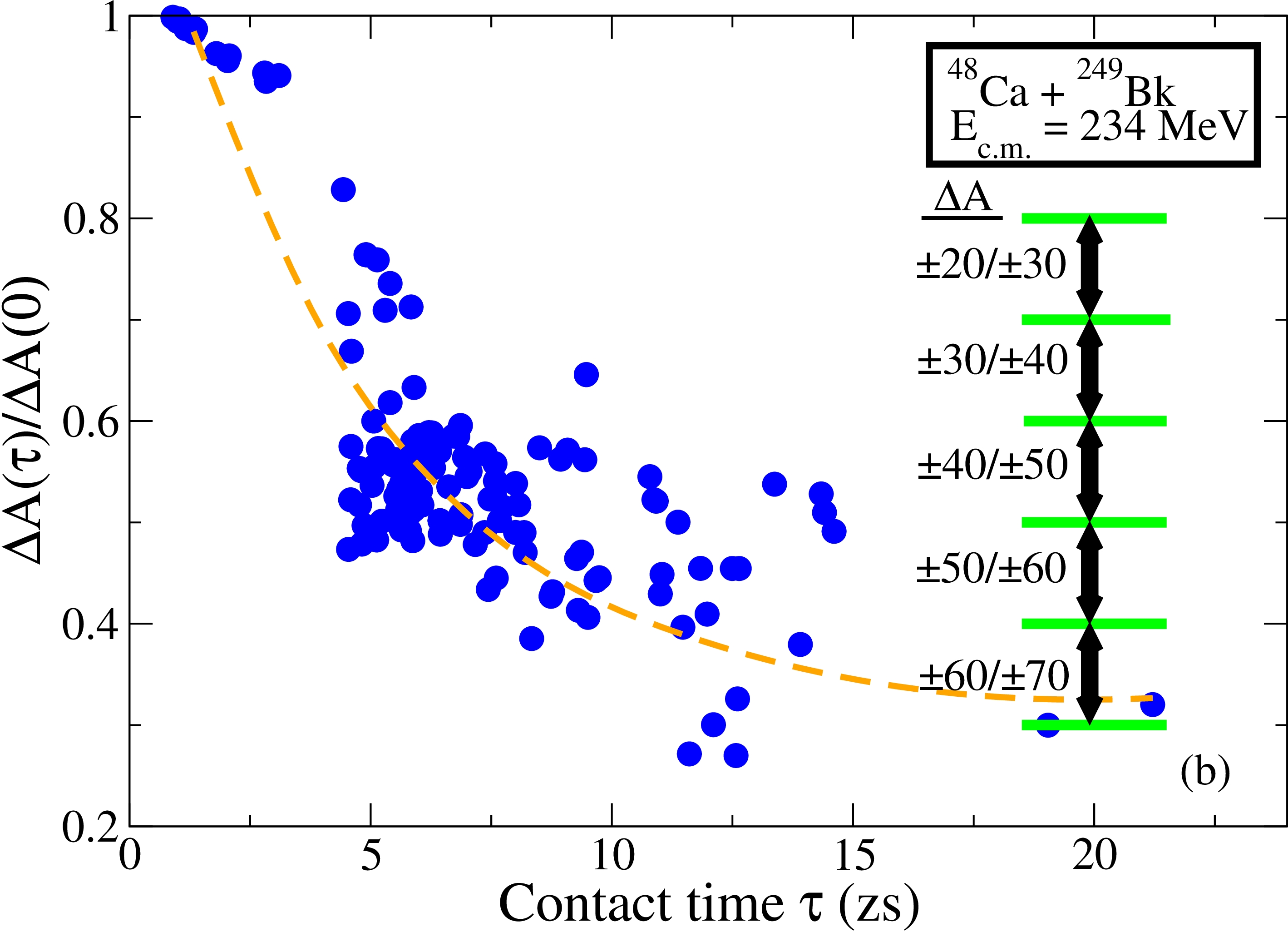}
\caption{(a) Mass ratio of fragment masses as a function of contact time
at $E_{\mathrm{c.m.}}=234$~MeV for the reaction $^{48}\mathrm{Ca}+{}^{249}\mathrm{Bk}$. The solid lines show possible fits.
(b) The ratio of the final and initial fragment masses as a function of contact time for the same reaction.
The dashed line shows one possible fit.\label{fig:MEQ}}
\end{figure}

Due to long reaction times, the quasifission process is also suitable to study the time-scale of mass equilibration.
Figure~\ref{fig:MEQ}(a) shows the mass ratio, $M_\mathrm{R}$, of fragment masses as a function of contact time $\tau$ (1~zs $ =10^{-21}$~s)
at $E_{\mathrm{c.m.}}=234$~MeV for the $^{48}$Ca+$^{249}$Bk reaction.
We define the contact time as the time interval between the time $t_1$
when the two nuclear surfaces
(defined as isodensities with half the saturation density $\rho_0/2=0.07$~fm$^{-3}$)
first merge into a single surface and
the time $t_2$ when the surface densities detach again.
The dashed line shows a characteristic fit of a function in the
form of $c_0+c_1exp(-\tau/\tau_0)$. Based on the quality of the fit and whether we exclude some
extreme points from the fit or not, we obtain equilibration
times between $8-10$~zs.
In Fig.~\ref{fig:MEQ}(b) we plot the ratio of final and initial mass difference between projectile-like fragment, $A_{PLF}$,
and target-like fragment, $A_{TLF}$, defined by, 
\begin{equation}
\Delta A(\tau) = A_{TLF}(\tau)-A_{PLF}(\tau)\;,
\label{dA}
\end{equation}
as a function of contact time $\tau$ for the
$^{48}\mathrm{Ca}+^{249}\mathrm{Bk}$ system at $E_{\mathrm{c.m.}}=234$~MeV.
The points correspond to the impact parameters used, ranging from head-on collisions to more peripheral
collisions and the full range of orientations angles for $^{249}\mathrm{Bk}$.
The horizontal lines on the right side of the figure indicate the net number of particles
transferred between the target and the projectile. We note that more mass transfer happens
at longer contact times as expected. From this figure we can also observe similar time-scale
for mass equilibration.
From these results (and others not shown here) we can conclude that mass equilibration takes substantially
longer in comparison to other quantities such as the equilibration of total kinetic energy (TKE) or $N/Z$ equilibration.
It is also interesting to observe that
there is clustering of results around certain mass ratios. This is shown to be related to shell effects influencing
the dynamical quasifission process in Ref.~\cite{godbey2019}.

\vspace{0.2in}
\subsection{Collective landscape}
\begin{figure*}[!htb]
        \includegraphics*[width=8.6cm]{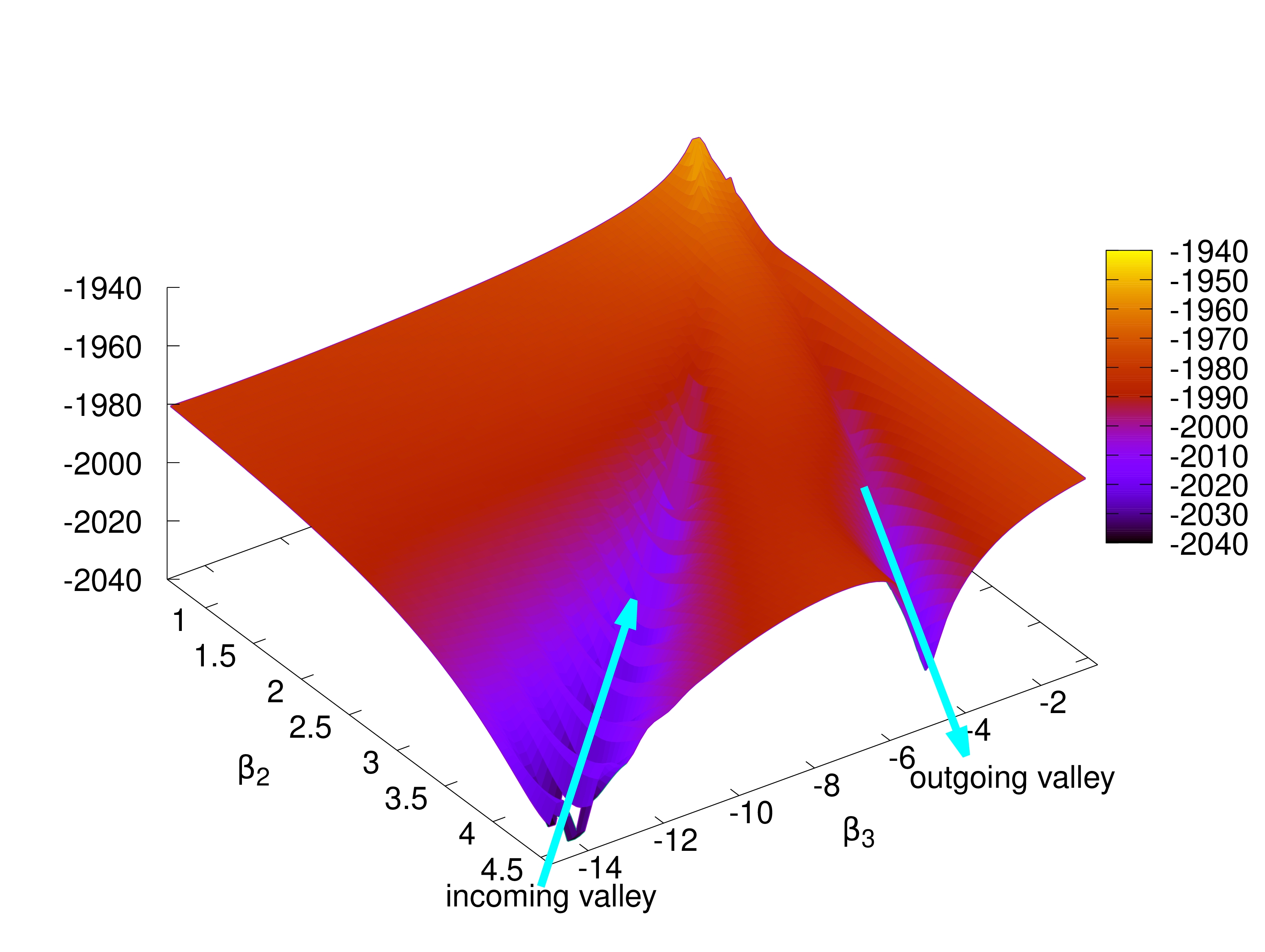}
        \includegraphics*[width=8.6cm]{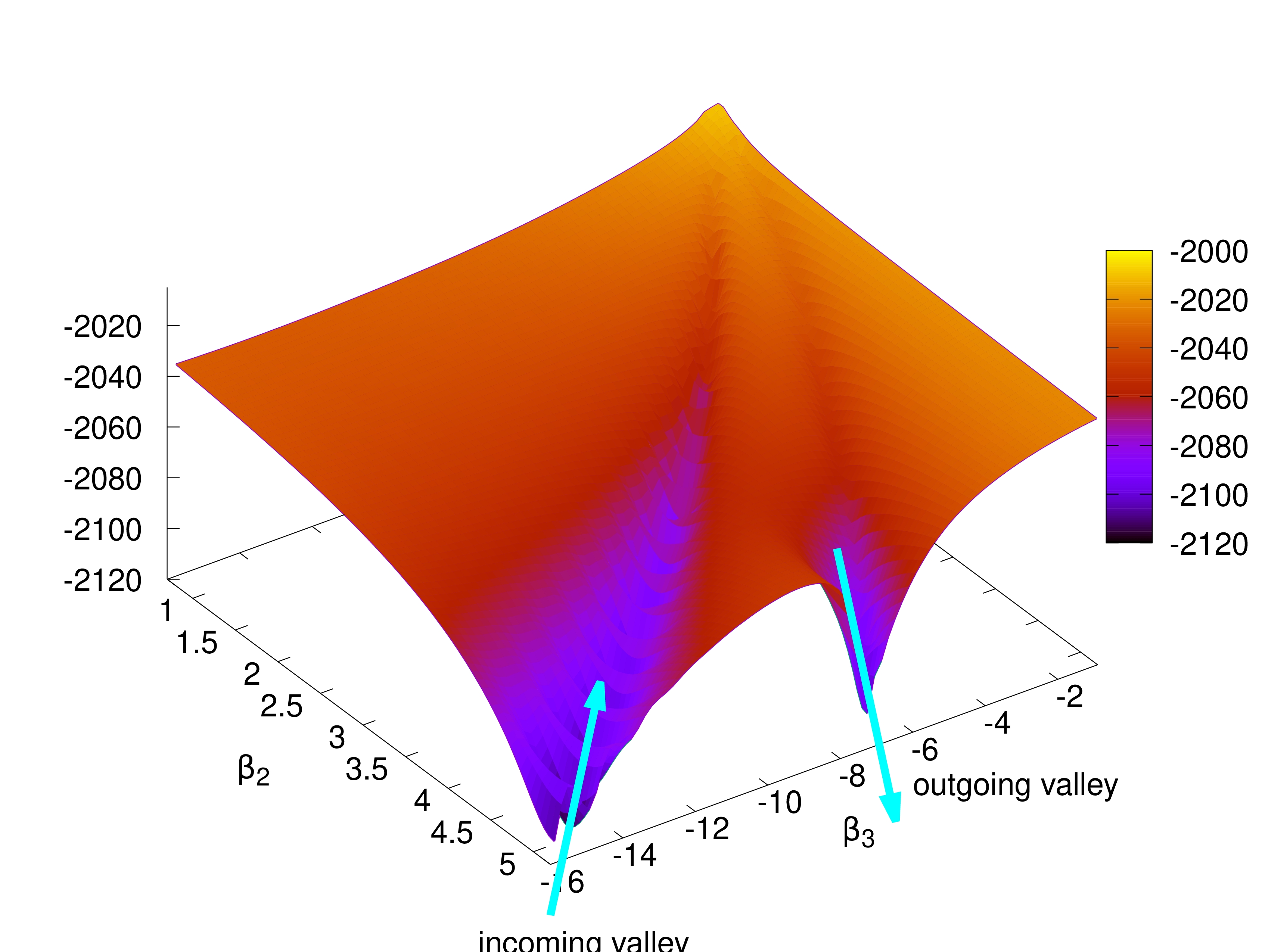}
        \caption{\protect
Plotted are the PESs calculated, using the DC-TDHF method, for central collisions of
$^{40,48}$Ca+$^{238}$U, with the equatorial orientation of
$^{238}$U. The PES on the left is for the
$^{40}$Ca+$^{238}$U system at $E_\mathrm{c.m.}=211$~MeV, while the PES on the
right is for the $^{48}$Ca+$^{238}$U system at $E_\mathrm{c.m.}=203$~MeV.
        }
        \label{fig:PES}
\end{figure*}

Quasifission and fusion-fission
could be used to help map out the non-adiabatic collective landscape between the fusion entrance channel
and the fission exit channel. It has been demonstrated that the TDHF theory is able to provide a good
simulation of the quasifission process. Calculated time-scales of quasifission indicate that while fast
quasifission events are dominant, much slower events resulting in a split with equal mass fragments have
also been observed. One of the open experimental questions is how to distinguish quasifission from
fusion-fission. This is important for the calculation of the evaporation residue formation
probability in superheavy element searches.
In Fig.~\ref{fig:PES} we show two such PESs calculated for the central collisions
of the
$^{40,48}$Ca+$^{238}$U systems, with the equatorial orientation of the
$^{238}$U. The PES on the left of Fig.~\ref{fig:PES} is for the
$^{40}$Ca+$^{238}$U system at $E_\mathrm{c.m.}=211$~MeV, while the PES on the
right is for the $^{48}$Ca+$^{238}$U system at $E_\mathrm{c.m.}=203$~MeV.
Surfaces in Fig.~\ref{fig:PES} are obtained by plotting the scattered $\beta_2$, $\beta_3$, and $E$
data obtained from the DC-TDHF calculations for the time-evolution of the nuclear density.
Since the scattered plot uses an extrapolation algorithm points far from the valleys may not be precise.
A number of observations can be made from the PESs shown in Fig.~\ref{fig:PES}. First,
we clearly see the valley corresponding to the incoming trajectory of the two nuclei.
As the system forms a composite the energy rises to maximum, but most likely never makes it
to the saddle point. The system spends a lot of time around this area undergoing complex
rearrangements and finally starts to proceed down the quasifission valley.

\vspace{0.2in}
\section{Summary}
Quasifission reactions have emerged as an interesting and vibrant area of research in recent years as they teach us about dynamical many-body effects at much longer time-scales compared to other heavy-ion reactions.
The persistence of shell effects for these time-scales has opened the possibility to view quasifission as a doorway process to fusion-fission and perhaps even fission.
This wide applicability positions quasifission as a vital process in understanding nuclear reactions across the board.
In advancing towards this goal, the TDHF theory and its extensions have emerged as an excellent theoretical tool to study these reactions.
The success of TDHF results in replicating experiment is particularly impressive as the calculations contain no free parameters.
Through the efforts of both theoretical and experimental study of quasifission, we have been able to identify a number of underlying physical phenomena affecting nuclear reactions, such as the dependence on mass-angle distributions on the orientation of deformed targets and the strong influence of shell effects in determination of reaction products.
These predictions take steps towards a more complete understanding of dynamical processes in nuclear reactions and may be crucial in determining such quantities as the $P_\mathrm{CN}$ by calibrating experimental angular distributions to that of the theory.
To this end methods and techniques to discern between quasifission and fusion-fission may emerge, paving the way for future studies of neutron-rich nuclei and superheavy elements.

\section*{Author Contributions}
The contribution of each of the authors has been
significant and the manuscript is the result of an even effort of both the authors. 

\section*{Funding}
This work was supported by the U.S. Department of Energy under grant No. DE-SC0013847.

\section*{Acknowledgments}

\bibliographystyle{frontiersinFPHY.bst}
\bibliography{VU_bibtex_master}
\end{document}